\documentclass[journal]{IEEEtran}
\ifCLASSINFOpdf
\else
   \usepackage[dvips]{graphicx}
\fi
\usepackage{url}

\usepackage{bm}
\usepackage{titlesec}
\usepackage{graphicx}
\usepackage{amsmath}
\usepackage{amsfonts}
\usepackage{multirow}
\usepackage{cite}
\begin{document}

\title{Novel Constructions of 2-D Golay Complementary Array Sets of Flexible Sizes}

\author{ Zhaoyu Zhang, Xiaoyu Chen,Yuqiong Du, and Luyi Zheng
\thanks{This work of Z. Zhang was supported in part by the National Natural Science Foundation of China under Grant 61601399, in part by Hebei Natural Science Foundation under Grant F2021203078, and in part by Science Research Project of  Hebei Education Department under Grant ZD2022026. }
\thanks{Zhaoyu Zhang, Xiaoyu Chen, Yuqiong Du and Luyi Zheng are with the School of Information Science and Engineer, Yanshan University, Qinhuangdao 066004, China, and also with the Hebei Key Laboratory of Information Transmission and Signal Processing, Qinhuangdao 066004, China (e-mail:  zhangzhaoyu176@163.com; chenxiaoyu@ysu.edu.cn; 17745797003@163.com; nanyoujiayu1999@163.com).}
}

\markboth{Journal of \LaTeX\ Class Files, Vol. XX, No. XX, XX 2024}
{Shell \MakeLowercase{\textit{et al.}}: Bare Demo of IEEEtran.cls for IEEE Journals}
\maketitle
\begin{abstract}
One-dimensional (1-D) Golay complementary sets (GCSs) possess numerous well-known properties and have achieved extensive use in communication engineering. The concept of 1-D GCSs can be extended to two-dimensional (2-D) Golay complementary array sets (GCASs). The letter proposes two constructions of 2-D GCASs based on two-dimensional extended generalized Boolean functions, where the set sizes and matrix parameters are flexible. 
The set size of GCASs with an array size of \(b^m \times b^n\) in the letter can reach \(N^{m+n+1}\), which has not been cited in the existing literature. 
Both structures are assembled directly and are not restricted by the parameters of other existing sequences, thus expanding the theoretical and applicative boundaries of Golay complementary array sets.
\end{abstract}

\IEEEpeerreviewmaketitle

\section{Introduction}

\IEEEPARstart{C}{omplementary} sequence pairs/sets have found many applications in science and engineering, particularly in wireless communication and signal processing. The concept of the one-dimensional Golay complementary sequence was first introduced by M. J. Golay in 1961 \cite{1}.  Subsequently, in \cite{2}, the concept of Golay complementary pairs was extended to Golay complementary sets with different set sizes. In 1999, Davis and Jedwab used generalized Boolean functions to construct binary Golay complementary pairs (GCPs) directly \cite{3}. In 2000, Paterson extended the results to the case of \(q\)-ary (where \(q\) is even) GCPs \cite{4}. The result, based on the generalized Boolean functions introduced in \cite{3} and \cite{4}, is referred to as the Golay-Davis-Jedwab (GDJ) pair, providing a significant contribution. Later, extensive research was conducted on several types of direct construction of Golay complementary sequences (GCSs) in references \cite{5,6,7,8 }, etc.

The GCPs and GCSs mentioned above can be extended to two-dimensional (2-D) arrays to broaden their application range, which are referred to as two-dimensional Golay complementary array pairs (GCAPs) \cite{9,10}, and two-dimensional Golay complementary array sets (GCASs) \cite{11,12}, respectively. GCAPs and GCASs have zero aperiodic autocorrelation sums at other positions besides the zero shift. GCASs can be used as precoding matrices \cite{13} in large-scale multi-input multi-output (MIMO) communication systems with uniform rectangular array (URA) configurations for omnidirectional transmission \cite{10,14}.

To date, the constructions of most 2-D GCAPs or GCASs still rely on indirect construction methods \cite{15,16}, which may be infeasible for quickly generating array sets. Pai and Chen proposed 2-D GCAPs and GCASs constructions based on the 2-D GBFs in \cite{9} and \cite{11}. In \cite{17}, direct constructions based on 2D-MVFs are presented. Later, Men proposed new GCASs structures in \cite{18}, which use two-dimensional extended generalized Boolean functions (which can be seen as a two-dimensional extension of 1-D GBFs  \cite{8}) to construct \(\left( {{N^k},{b^{{m_1}}},{b^{{m_2}}}} \right)\)-GCASs and \(\left( {{{N}}_1^{k_1}{{N}}_2^{{k_2}},b_1^{{m}},b_2^{{n}}} \right) \)-GCASs. Recently, in \cite{12}, by truncating certain arrays generated by the 2-D EGBFs, new two-dimensional GCAS constructions are proposed. 


The direct construction offers high flexibility, enabling the generation of more diverse and adaptable sequences. 
In particular, the proposed constructions based on 2-D EGBFs possess specific algebraic properties and are not constrained by the parameters of other existing sequences, allowing for more efficient design of sequences to meet specific requirements.
In the letter, we propose two direct constructions of two-dimensional GCASs with new flexible parameters based on 2-D EGBFs . The resulting GCASs provide a greater number of arrays, which has not been mentioned in the literature. 

The remainder of the letter is organized as follows. In Section II, we introduce the notation used in the letter, briefly discuss EGBFs, and define the concepts of AACFs and GCASs. In Section III, we present two new structures of GCASs based on 2-D EGBFs. Finally, Section IV concludes the letter.

\section{Preliminaries and Notations}
The following notations will be employed throughout the letter:

• \({\left(  \cdot  \right)^*}\) denotes the complex conjugation. 

• \({\left(  \cdot  \right)^T}\) denotes the transpose. 

• For an integer \(b \ge 2\), \(U\left( b \right)\) denotes the set of integers \(b\) and \(r\), satisfying \({gcd}\left( {{b,r}} \right){= 1}\), where \(r\) takes values from \(1\) to \(b - 1\).

• \({lcm}\left(  \cdot  \right)\) refers to the least common multiple.

• \({\mathbb{Z}_q}=\left\{ {0,1, \ldots ,q - 1} \right\}\) is the ring of integers modulo \(q\).

• \({\left( A \right)_{g,i}}\) denotes the \(\left( {g,i} \right)\)-th element of the matrix \(A\).

• \(\xi_q  = {e^{2\pi \sqrt { - 1} /q}}\), \({\psi _q}\left( {{f_{g,i}}} \right)\) is denoted by \(\xi _q^{{f_{g,i}}}\). 

\subsection{2-D Golay Complementary Arrays Sets}
{\em Definition 1 (2-D aperiodic autocorrelation function):} The 2-D aperiodic autocorrelation function of \(C\) at shift \(({{u_1},{u_2}})\) is denoted by \(\rho \left( {C;{u_1},{u_2}} \right)\), whose formula is presented as follows:  
\begin{align}
\rho \left( C; {u_1}, {u_2} \right) = &\sum\limits_{g = 0}^{{L_1} - 1} \sum\limits_{i = 0}^{{L_2} - 1} {C_{g + {u_1}, i + {u_2}}} C_{g, i}^* \notag\\
&\text{for } - {L_1} < {u_1} < {L_1} \text{ and } - {L_2} < {u_2} < {L_2},
\end{align}
where \({C_{g + {u_1}, i + {u_2}}=0}\) for \(\left( {g + {u_1}} \right) \notin \left\{ {0,1, \ldots ,{L_1} - 1} \right\}\) or \(\left( {i + {u_2}} \right) \notin \left\{ {0,1, \ldots ,{L_2} - 1} \right\}\).  Note that \(\rho \left( {C;{u_1},{u_2}} \right) = {\rho ^*}\left( {C; - {u_1},{u_2}} \right)\).

{\em Definition 2 (Golay Complementary Array Set):} The set of arrays \(C = \left\{ {{C_l}|l = 0, \ldots ,N - 1} \right\}\), where each array \({C_l}\) is of size \({L_1} \times {L_2}\). The set \(C\) is referred to as an \(\left( {N,{L_1},{L_2}} \right) - \)GCAS if 
\begin{align}
\sum\limits_{l = 0}^{N - 1} {\rho \left( {{C_l};{u_1},{u_2}} \right)}  = \left\{ {\begin{array}{*{20}{c}}
{N{L_1}{L_2},\left( {{u_1},{u_2}} \right) = \left( {0,0} \right)}\\
{0,\hspace{1cm}\left( {{u_1},{u_2}} \right) \ne \left( {0,0} \right)}.
\end{array}} \right. 
\end{align}
If \({C_l}=\left( {{\xi ^{{c_l}}}} \right)\) where \({c_l}\) is \(q\)-ary array over \({\mathbb{Z}_q}\) for \(l = 0,1,\ldots, N-1\), we define the array set \(\left\{ {{c_0},{c_1}, \ldots ,{c_{N - 1}}} \right\}\) a \(q\)-ary GCAS. Indeed, a GCAP is a \(\left( {2,{L_1},{L_2}} \right) - \)GCAS. 
\subsection{2-D Extend Generalized Boolean Functions}
A 2-D EGBF \(f\): \({\mathbb{Z}_{{b_1}}^m }\times {\mathbb{Z}_{{b_2}}^n} \to {\mathbb{Z}_q}\)  is a function mapping in \(m + n\) \(q\)-ary variables \({x_1},{x_2}, \ldots ,{x_m},{y_1},{y_2}, \ldots ,{y_n}\). Let \(2 \le {b_1},{b_2} \le q\) , \({x_i} \in \mathbb{Z}_{b_1} \) , \({y_j} \in \mathbb{Z}_{b_2} \), \( {i \in \left[ {1,m} \right]} \), \( {j \in \left[ {1,n} \right]} \) and \(f\left( {\left( {{x_1},{x_2}, \ldots ,{x_m}} \right),\left( {{y_1},{y_2}, \ldots ,{y_n}} \right)} \right) \in \mathbb{Z}_{q}\). For a 2-D EGBF with \(m + n\) \(q\)-ary variables, the 2-D \(\mathbb{Z}_{q}\)-valued array is defined as
\begin{equation}
F = \left[ {\begin{array}{*{20}{c}}
{{f_{0,0}}}&{{f_{0,0}}}& \cdots &{{f_{0,b_2^n - 1}}}\\
{{f_{1,0}}}&{{f_{1,1}}}& \cdots &{{f_{1,b_2^n - 1}}}\\
 \vdots & \vdots & \ddots & \vdots \\
{{f_{b_1^m - 1,0}}}&{{f_{b_1^m - 1,1}}}& \cdots &{{f_{b_1^m - 1,b_2^n - 1}}}
\end{array}} \right]. 
\end{equation} 
Through the function, the array we can obtain is of size of \(b_1^m \times b_2^n\), which is given by setting \({f_{g, i}} = f\left( {\left( {{g_1},{g_2}, \ldots,{g_m}} \right),\left( {{i_1},{i_2}, \ldots,{i_n}} \right)} \right)\). Let \(\left( {{g_1},{g_2}, \ldots ,{g_m}} \right)\) and \(\left( {{i_1},{i_2}, \ldots ,{i_n}} \right)\) be \(b_1\)-ary and \(b_2\)-ary vector representations of integers \(g = \sum\nolimits_{\gamma  = 1}^m {{g_\gamma }b_1^{\gamma  - 1}} \) \(\left({g_\gamma}\in {\mathbb{Z}_{b_1}}\right )\) and \(i = \sum\nolimits_{\gamma  = 1}^n {{i_\gamma }b_2^{\gamma  - 1}} \) \(\left({i_\gamma}\in {\mathbb{Z}_{b_2}}\right ) \), respectively. For convenience, we abbreviate the complex-valued array as \(\bm{f} = {\psi _q}\left( F \right)\).

\section{GCASs Based on EGBFs}
\label{sec:guidelines}
\subsection{Proposed Constructions}

{\em Theorem 1:} Consider nonnegative integers \(n,m,b\) and \(k\) with \(b \ge 2\) and \(1 \le k \le m + n\). Let nonempty sets \({I_1},{I_2}, \ldots ,{I_k}\) be a partition of \(\left\{ {1,2, \ldots ,m + n} \right\}\). \({\pi _\alpha }\) is defined as a bijection from \(\left\{ {1,2, \ldots ,t _\alpha} \right\}\) to \({T _\alpha }\) for \(\alpha  = 1,2, \ldots ,k\), where \({t_\alpha } = \left| {{I_\alpha }} \right|\). Define a 2-D EGBF \(f\) :\(\mathbb{Z}_b^m \times \mathbb{Z}_b^n \to {\mathbb{Z}_q}\) as follows: 

\begin{equation}
f=\frac{q}{b}\sum_{\alpha =1}^{k}\sum_{\beta =1}^{t_{\alpha } -1} d_{\alpha ,\beta }  z_{\pi _{\alpha }(\beta ) }z_{\pi _{\alpha }(\beta+1 ) } +\sum_{\gamma =1}^{q-1}\sum_{l =1}^{m+n}\lambda _{\gamma ,l}z_{\gamma }^{l}+\lambda _{0} 
\end{equation}
where \({z_l} \in\mathbb{Z} _{b} \), \({z_1} = {x_1},{z_2} = {x_2}, \ldots ,{z_m} = {x_m},{z_{m + 1}} = {y_1}, \ldots ,{z_{m + n}} = {y_n}\), \(b\left| q \right.\), \({d_{\alpha ,\beta }} \in U\left( b \right)\), and \({\lambda _{\gamma ,l}},{\lambda _0} \in \mathbb{Z} _{q}\). Consider a positive integer \(N\) satisfying \(N\left| q \right.\) and \(N \ge b\). Let \({n_\alpha } \in {\mathbb{Z}_N}\), then the array set

\begin{equation}
\mathcal{A}=\left \{ f+\frac{q}{N}\left ( \sum_{\alpha =1}^{k}n_{\alpha}  z_{\pi _{\alpha }(1 ) } +n_{k+1}  z_{m}\right )    \right \} 
\end{equation}
forms a \(q\)-ary \(\left( {{N^{k + 1}},{b^{m}},{b^{n}}} \right)\)-GCAS. 

{\em Proof:} The proof can be found in the Appendix A. 

{\em Example 1:} Let \(N = 3\), \(q = 6\), \(b=2\), \(m = 1\), \(n = 3\) and \(k = 1\) in Theorem 1, taking \({\pi _1} = \{ 4,1,2,3\} \). Then we can get a 2-D EGBF \(f = 3{z_4}{z_1} + 3{z_1}{z_2} + 3{z_2}{z_3}=3{y_3}{x_1}+3{y_1}{x_1}+3{y_1}{y_2}\). Through the function we can obtain a \(\left( {9,2,8} \right) \)- GCAS, denoted as \({\cal A} = \left\{ {f + 2{n_1}{y_3} + 2{n_2}{x_1},{n_1},{n_2} \in {\mathbb{Z}_3}} \right\}\). The set \(\mathcal{A}\) is listed in Table I. Following the example, a GCAS with array size \({2^{m}} \times {2^n}\) can be generated, and the number of sets is 9. 
\renewcommand{\arraystretch}{1.8}  
\begin{table}[h]
\caption{ \(\mathcal{A}\) Derived From Example 1}
\centering
\begin{tabular}{|c|c|c|}  
\hline
\multicolumn{3}{|c|}{\textbf{\(\left( {9,2,8} \right)-\){GCAS} \(\mathcal{A}\)}}  \\  
\hline
\({\mathcal{A}}^0=\left[ {\begin{array}{*{20}{c}}{00030003}\\{03003033}\end{array}} \right] \)
& \({\mathcal{A}}^1=\left[ {\begin{array}{*{20}{c}}{00030003}\\{25225255}\end{array}} \right]\) 
& \({\mathcal{A}}^2=\left[ {\begin{array}{*{20}{c}}{00032225}\\{03005255}\end{array}} \right] \) \\
\hline
\({\mathcal{A}}^3=\left[ {\begin{array}{*{20}{c}}{00032225}\\{25221411}\end{array}} \right] \)
& \({\mathcal{A}}^4=\left[ {\begin{array}{*{20}{c}}{00032225}\\{41443033}\end{array}} \right]\) 
&\( {\mathcal{A}}^5=\left[ {\begin{array}{*{20}{c}}{00030003}\\{41441411}\end{array}} \right] \) \\
\hline
\({\mathcal{A}}^6=\left[ {\begin{array}{*{20}{c}}{00034441}\\{03001411}\end{array}} \right]\) 
& \({\mathcal{A}}^7=\left[ {\begin{array}{*{20}{c}}{00034441}\\{25223033}\end{array}} \right]\) 
& \({\mathcal{A}}^8=\left[ {\begin{array}{*{20}{c}}{00034441}\\{41445255}\end{array}} \right] \) \\
\hline
\end{tabular}
\end{table}

{\em Remark 1: }Note that when \(k\) takes the maximum, i.e., \(k = m+n\), then we can get a \(\left( {{N^{m+n+1}},{b^m},{b^n}} \right)\)-GCAS, which has not been reported in previous literature.  
In the following, Theorem 1 is further extended to construct GCASs with various array set parameters.

{\em Theorem 2: }Consider nonnegative integers \( n\), \(m\), \(b_1\), \(b_2\), \(k_1\) and \(k_2\) with \(b_1,b_2\ge 2\) and \(m,n\ge 1\). Let nonempty sets \(I_1,I_2, \ldots ,I_{k_1}\) be a partition of \(\left\{ 1,2, \ldots ,m \right\}\) and nonempty sets \(I_1^{\prime}\), \(I_2^{\prime}\), \ldots , \(I_{k_2}^{\prime}\) be a partition of \(\left\{ 1,2, \ldots ,n \right\}\), where \(k_1 \le m\), \(k_2 \le n\). Take \({t_\alpha }\) and \(t_\mu ^{\prime}\) as orders of \({I_\alpha}\) and \({I_\mu ^{\prime}}\), respectively. Let \({\pi _\alpha }\) be a bijection mapping from \({\left\{ {1,2, \ldots ,{t_\alpha }} \right\}}\) to \({I_\alpha }\) and \({\sigma _\mu }\) be a bijection from \({\left\{ {1,2, \ldots ,t_\mu ^{\prime}} \right\}}\) to \({I_\mu ^{\prime}}\), where \({\alpha  = 1,2, \ldots ,k_1}\) and \({\mu  = 1,2, \ldots ,k_2}\). The 2-D EGBF \(f\) : \({\mathbb{Z}_{b_1}^m \times \mathbb{Z}_{b_2}^n \to {\mathbb{Z}_q}}\) is defined as: 
\begin{multline}
f = \frac{q}{b_{1}} \sum_{\alpha =1}^{k_{1}} \sum_{\beta =1}^{t_{\alpha} -1} d_{\alpha ,\beta} x_{\pi_{\alpha}(\beta)} x_{\pi_{\alpha}(\beta+1)} + \sum_{\gamma =1}^{q-1} \sum_{l =1}^{m_{1}} \lambda_{\gamma, l} x_{l}^{\gamma} + \lambda_{0} \\
 + \frac{q}{b_{2}} \sum_{\mu =1}^{k_{2}} \sum_{\nu =1}^{t_{\mu}^{'}-1} d_{\mu, \nu}^{'} y_{\sigma_{\mu}(\nu)} y_{\sigma_{\mu}(\nu +1)} 
 + \sum_{\gamma =1}^{q-1} \sum_{l =1}^{m_{2}} \nu_{\gamma, l} y_{l}^{\gamma}
\end{multline}
where \({x_l} \in \mathbb{Z}_{b_1} \), \({y_l} \in \mathbb{Z}_{b_2} \),\(\left. {{b_1}} \right|q\), \(\left. {{b_2}} \right|q\), \(d_{\alpha ,\beta}\in U\left( {{b_1}} \right) \), \({d_{\mu ,\upsilon }^{\prime}} \in U\left( {{b_2}} \right)\) and \({\lambda _{\gamma ,l}},{\upsilon _{\gamma ,l}},{\lambda _0} \in \mathbb{Z}_q \). If there exist two positive integers \(N_1\) and \(N_2\) satisfying \(\left. {{N_1}} \right|q, {{N_2}} |q\) , \({N_1}  \ge  {b_1}\), \({N_2} \ge {b_2}\) and \({n_\alpha }\), \({n_{{k_1}+1} }\in {\mathbb{Z}_{N_1}}\), \({n_\mu ^{\prime}}\in {\mathbb{Z}_{N_2}}\), the array set
\begin{multline}
 \mathcal{A}=\left. \Bigg\{ f+\frac{q}{N_{1} }\sum_{\alpha =1}^{k_{1} }n_{\alpha}  x_{\pi _{\alpha }(1 ) } +\frac{q}{N_{2} }\sum_{\mu  =1}^{k_{2} }n_{\mu } ^{\prime}  y_{\sigma _{\mu }(1 ) } \right.
\\  \left.  +n_{k_{1} +1}  \left (x_{k_{1} +1}+y_{k_{1} +1}  \right ) \right. \Bigg\}
\end{multline}
is a \(q\)-ary \(\left( {{{N}}_1^{{k_1} + 1}{{N}}_2^{{k_2}},b_1^{{m}},b_2^{{n}}} \right) \)- GCAS

{\em Proof: }The proof can be found in the Appendix B.  

{\em Remark 2:}  
In contrast to Theorem 1, GCASs with more flexible array sizes have been obtained from Theorem 2, taking the form \(b_1^m \times b_2^n\), which provides enhanced flexibility in array design.

{\em Remark 3:}
Let \(k_1\) and \(k_2\) both be set to their maximum values, i.e., \({k_1}+{k_2}=m+n\), we can get a \(\left( {{{N}}_1^{m + 1}{N}_2^{n},b_1^{{m}},b_2^{{n}}} \right) \)- GCAS, which has larger array number than previous literature under the same conditions.

\subsection{Comparison With the Previous Works}

We have listed the known GCASs for comparison in Table II, and the advantages of our proposed constructions are as follows:

Compared with \cite{11} and \cite{17}, more flexible-sized GCASs can be obtained by setting the generalized variables \(N\) and \(b\). The constructions we propose can provide more array quantities in one set, which is not mentioned before. 
\renewcommand{\arraystretch}{2.5}  
\begin{table}[t]
\centering
\caption{Comparison of Parameters GCASs With Others}
\begin{tabular}{|p{1.1cm}|p{3.3cm}|p{1.6cm}|p{1cm}|}

\hline
Ref.           & \parbox{3.5cm}{\centering Parameters }  & Phase \(q\)          & Method                   \\[0ex] \hline

\cite{11} & \parbox{3.5cm}{\centering \( (2^k, 2^{m_1}, 2^{m_2}) \) \\ \( m_1 + m_2 \geq k \geq 1 \)} & \parbox{1.2cm}{\( q \ge 2 \) and \(q\) is even} & \parbox{0.9cm}{2-D GBFs} \\[1ex] \hline
Th1.\cite{17} & \parbox{3.5cm}{\centering \( \left( p_1^{k_1} p_2^{k_2}, p_1^{m_1}, p_2^{m_2} \right) \) \\ \( m_1 \geq k_1 \geq 1 \), \( m_2 \geq k_2 \geq 1 \)} & \parbox{1.2cm}{\(lcm\left( {{p_1},{p_2}} \right)|q\)} & \multirow{2}{*}{\parbox{0.9cm}{2-D MVFs}} \\[1.8ex] \cline{1-3}
Th2.\cite{17} & \parbox{3.5cm}{\centering \( \left( p^k, p^{m_1}, p^{m_2} \right) \) \\ \( m_1 + m_2 \geq k \geq 1 \)} & \( p|q \) & \\[1ex] \hline
Th5.\cite{16} & \parbox{3.5cm}{\centering \( (N^k, N^m, N^n), {m,n\ge 1} \)} & \( q = N \) & seed PU matrices \\[-1ex] \hline
Th1.\cite{18} & \parbox{3.5cm}{\centering \( \left( {N^k}, {b^m}, {b^n} \right)\)\\\( N \ge b \ge 2 ,m + n \ge k \ge 1 \) }& \(lcm\left( N,b \right) | q \) & \multirow{2}{*}{\parbox{0.9cm}{\centering 2-D EGBFs}} \\[1ex] \cline{1-3}
Th2.\cite{18}& \parbox{3.5cm}{\centering \( \left( N_1^{k_1} N_2^{k_2}, b_1^m, b_2^n \right)\) \\ \(m \ge k_1 \ge 1, n \ge k_2 \ge 1\) \\\(  N_1 \ge b_1 \ge 2, N_2 \ge b_2 \ge 2 \)} & \( \sigma \mid q \)&  \\[2.5ex] \hline
Th1.\cite{12} &\parbox{3.5cm}{\centering\( \left( {N^k},b^{m},b^{n-1}+{\eta _1} \right) \) \( N \ge b \ge 2 ,m + n \ge k \ge 1 \) }& \(lcm\left( N,b \right) | q \)  & \multirow{2}{*}{\parbox{0.9cm}{2-D EGBFs}} \\[1ex] \cline{1-3}
Th2.\cite{12} & \parbox{3.5cm}{\centering\(\left( {{{N}}_1^{k_1} }{{N}}_2^{{k_2}},b_1^{m},b_2^{n-1}+\eta _2 \right) \)\\ \(m \ge k_1 \ge 1, n \ge k_2 \ge 1\) \\\(  N_1 \ge b_1 \ge 2, N_2 \ge b_2 \ge 2 \)} & \( \sigma \mid q \)&\\[2.5ex] \hline
Th.1   &\parbox{3.5cm}{\centering\( \left( {N^{k+1}},b^{m},b^{n} \right) \) \( N \ge b \ge 2 ,m + n \ge k \ge 1 \) }& \(lcm\left( N,b \right) | q \)  & \multirow{2}{*}{\parbox{0.9cm}{2-D EGBFs}} \\[1ex] \cline{1-3}
Th.2 & \parbox{3.5cm}{\centering\(\left( {{{N}}_1^{{k_1} + 1}{{N}}_2^{{k_2}},b_1^{{m}},b_2^{{n}}} \right) \), \\ \(m \ge k_1 \ge 1, n \ge k_2 \ge 1\) \\\(  N_1 \ge b_1 \ge 2, N_2 \ge b_2 \ge 2 \)} & \( \sigma \mid q \) &\\[2.5ex] \hline
\end{tabular}
\raggedright 
\(\sigma  = lcm\left( {{N_1},{N_2},{b_1},{b_2}} \right)\)

\(  {\eta _1} = \sum\nolimits_{\alpha  = 1}^{k - 1} {{r_\alpha } \cdot {b^{n - k + \alpha  - 1}} + {r_0} \cdot {b^{n - k}}} ,{r_\alpha },{r_0} \in \mathbb{Z}_b \)

\(  {\eta _2} = \sum\nolimits_{\alpha  = 1}^{{k_2} - 1} {r_\alpha } \cdot {b_2^{\pi _2\left(n - k_2 + \alpha \right) - 1} + {r_0} \cdot {b_2^\varphi }} ,{r_\alpha },{r_0} \in \mathbb{Z}_{b_2},0  \le \varphi \le {n-{k_2}}\)
\end{table}
The maximum number of arrays generated by \cite{12,18} is \(N^{m+n}\), while the maximum number of arrays that can be obtained in Theorem 1 is \(N^{m+n+1}\).  Under the same conditions as those in Example 1, which presents the \(\left(9,2,8\right)\)-GCAS \(\mathcal{A}\), while the \(\left(3,2,8\right)\)-GCAS \(\mathcal{B}\) is obtained by using the method in Theorem 1 of \cite{18}. To demonstrate the differences between the two constructions clearly, the algebraic expressions of Example I are listed in Table III.
New parameter GCASs with larger set sizes can be obtained by the constructions, such as (\(3^4,2^1,2^2\))-GCASs, (\(4^{12},3^5,3^6\))-GCASs, (\(5^8,4^5,4^2\))-GCASs and so on, which cannot be obtained by other constructions. Due to limited space, we will not list them one by one.
\renewcommand{\arraystretch}{1.8}  
\begin{table}[h]
\centering
\caption{Algebraic Expressions of Different Constructions}
\begin{tabular}{|c|c|}  
\hline
\( \mathcal{A}\)
&\(\left\{ f + 2{n_1}{y_3} + 2{n_2}{x_1},n_1,n_2 \in \mathbb{Z}_3 \right\}\)\\
&=\(\left\{\begin{array}{cc}f,f+2{y_3},f+4{y_3},f+x_1,f+2{x_1},f+2{y_3}+x_1,
\\f+2{y_3}+2{x_1},f+4{y_3}+{x_1},f+4{y_3}+2{x_1}\end{array}\right\}  \)\\
\hline
\( \mathcal{B}\)
&\(\left\{ f + 2{n_1}{y_3},n_1\in \mathbb{Z}_3 \right\}\)
\\&\(=\left\{\begin{array}{cc}f,f+2{y_3},f+4{y_3}\end{array} \right\} \) \\
\hline
\end{tabular}
\end{table}

The corresponding functions of the multi-dimensional GCASs are extracted from the high-dimensional seed PU matrixs, and the 2-D GCASs are obtained through projection in \cite{16}. The constructions in the letter can directly obtain GCASs by using two-dimensional EGBFs without the need for various matrix operations, thus ensuring higher efficiency. Consequently, the constructions we propose can quickly generate array sets with a larger number of arrays, enabling the creation of diverse and flexible sequences, and offering more possibilities for applications in the field of communication.

\section{Conclusion}
In this letter, we propose two direct constructions of two-dimensional GCASs with noval parameters based on the two-dimensional EGBFs. The proposed structures can generate \(\left( {{{{N}}^{k + 1}},{b^{m}},{b^{n}}} \right)\)-GCAS and \(\left( {{{N}}_1^{{k_1} + 1}{{N}}_2^{{k_2}},b_1^{{m_1}},b_2^{{m_2}}} \right) \)- GCAS, both of which offer highly flexible array sizes. The proposed GCASs contain more complementary Golay arrays, which have not been reported yet. This provides greater flexibility for sequence design in practical communication systems. Furthermore, the resulting GCASs can be directly derived from the two-dimensional extended generalized Boolean functions without the need for special two-dimensional arrays or specific one-dimensional sequences.


\renewcommand\appendix{\par
\setcounter{section}{0}
\setcounter{subsection}{0}
\gdef\thesection{Appendix \Alph{section}}}
\appendix

\section*{Appendix A}
\section*{Proof of theorem 1}

\textit{Proof:} First, we define \(h = g + {u_1}\), \(j = i + {u_2}\). Set a \(N\)-ary vector \(\bm{\mu}=[\mu_1,\mu_2,…,\mu_k]\). Let \(\left( {{g_1},{g_2}, \ldots ,{g_m}} \right)\), \(\left( {{h_1},{h_2}, \ldots ,{h_m}} \right)\), \(\left( {{i_1},{i_2}, \ldots ,{i_n}} \right)\) and \(\left( {{j_1},{j_2},\ldots ,{j_n}} \right)\) be the \(b\)-ary representation of integers \(g\), \(h\), \(j\) and \(i\), respectively. 

To prove the \(\mathcal{A}\) is GCAS, for any \( c \in \mathcal{A}\), we need to demonstrate that 
\begin{multline}
\sum\limits_{c \in \mathcal{A}} {\sum\limits_{g = 0}^{{b^m} - 1} {\sum\limits_{i = 0}^{{b^n} - 1} {{\xi ^{{c_{g + {u_1},i + {u_2}}} - {c_{g,i}}}}} } }  \\
= \sum\limits_{g = 0}^{{b^m} - 1} {\sum\limits_{i = 0}^{{b^n} - 1} {\sum\limits_{\bm{\mu}} {{\xi ^{{c^\mu_{g + {u_1},i + {u_2}}} - {c^\mu_{g,i}}}}} } }  = 0,
\end{multline}
where \(\left( {{u_1},{u_2}} \right) \ne \left( {0,0} \right)\). The \(b\)-ary representations of \(g\), \(h\), \(j\) and \(i\) are given as follows:
\begin{align}
\begin{array}{l}
{a_l} = \left\{ \begin{array}{l}
{g_l}\hspace{1cm}for\hspace{0.1cm}1 \le l \le m;\\
{i_{l - n}}\hspace{0.6cm}for\hspace{0.1cm} m < l \le m + n,
\end{array} \right.\\
{b_l} = \left\{ \begin{array}{l}
{h_l}\hspace{1cm}for\hspace{0.1cm}1 \le l \le m;\\
{j_{l - n}}\hspace{0.6cm}for\hspace{0.1cm} m < l \le m + n,
\end{array} \right.
\end{array}
\label{}
\end{align}
In what follows, the proof of Theorem 1 will be taken into account from two cases. 

{\em Case 1:} In the case, we have \({a_{{\pi _\alpha }\left( 1 \right)}} \ne {b_{{\pi _\alpha }\left( 1 \right)}}\) for \(\alpha =1,2, \ldots ,k \) or \(a_m\ne b_m\). Set \(\pi^{\prime}(\alpha)={\pi _\alpha }\left( 1 \right)\) for \(\alpha =1,2, \ldots ,k \), \(\pi^{\prime}(k+1)=m\). Then  \(\pi^{\prime}(\alpha) \) is a bijective mapping from \(\mathbb{N} _{k+1}\) to \(\mathbb{N} _m\). Due to \(b \le N\), \(\bm{\mu}\cdot (a_{{\pi _\alpha }\left( 1 \right)} - b_{\pi _\alpha(1)} )\) mod \(N\) or \(\bm{\mu}\cdot (a_m -b_m )\) mod \(N\) takes the values \(0\) to \(N-1\) equally always for the condition, implying that \(\textstyle \sum_{\bm{\mu}} {\xi ^{{\bm{\mu}}\cdot(a_{\pi ^{\prime}_\alpha }-b_{\pi ^{\prime}_\alpha })} } = 0\). 
Therefore, we have
\begin{equation}
    \sum\limits_{c \in \mathcal{A}} {{\xi ^{{c_{h,j}} - {c_{g,i}}}}}  = 0\text{.} 
\end{equation}

{\em Case 2:}In that case, we have \({a_{{\pi _\alpha }\left( 1 \right)}} = {b_{{\pi _\alpha }\left( 1 \right)}}\) for all \(\alpha  =1,2, \ldots ,k\). We assume \({a_{{\pi _\alpha }\left( \beta  \right)}} = {b_{{\pi _\alpha }\left( \beta \right)}}\) for  \(\alpha  =1,2, \ldots ,\hat \alpha  - 1\) and \(\beta=1,2, \ldots ,{m_\alpha }\). In addition, we assume that \(\hat \beta\) is the smallest number satisfying \({a_{\pi_{\hat{\alpha}} (\hat{\beta})}} \ne {b_{\pi_{\hat{\alpha}} (\hat{\beta})}}\). The integers \({a^s}\) and \({b^s}\) are only different with \(a\) and \(b\) in the position, respectively, where \(s=1,2,…,b-1\). Let \(a_{\pi_{\hat{\alpha}}(\hat{\beta} - 1)}^s= 1 - a_{\pi_{\hat{\alpha}}(\hat{\beta} - 1)}\) and \(b_{\pi_{\hat{\alpha}}(\hat{\beta} - 1)}^s = 1 - b_{\pi_{\hat{\alpha}}(\hat{\beta} - 1)}\). 
If \(1 \le {\pi _{\hat \alpha }}\left( {\hat \beta \; - 1} \right) \le m\), we have
\begin{equation}
\begin{array}{l}
c_{{g^s,i}} - c_{g,i} \\
\equiv s\dfrac{q}{b}\left( {d_{\hat{\beta}-2}{a_{{\pi _{\hat \alpha }}(\hat \beta  - 2)}} + d_{\hat{\beta}}{a_{{\pi _{\hat \alpha }}(\hat \beta )}}} \right)\left( {1 - 2{g_{{\pi _{\hat \alpha }}(\hat \beta  - 1)}}} \right) \\
 + {p_{{\pi _{\hat \alpha }}(\hat \beta  - 1)}}\left( {1 - 2{g_{{\pi _{\hat \alpha }}(\hat \beta  - 1)}}} \right)\text{ \hspace{0.1cm} (mod {\em q}).} 
\end{array}
\end{equation}
where \(a_{\pi (t - 1)}^s = g_{\pi (t - 1)}^s\) and \(a_{\pi (t - 1)}= g_{\pi (t - 1)}\). Since  \(a_{\pi (t - 2)}= b_{\pi (t - 2)}\) and  \(a_{\pi (t - 1)}= b_{\pi (t - 1)}\), from (9), we can obtain
\begin{equation}
\begin{aligned}
{c_{{h^s},j}} - {c_{h,i}} &\equiv s\dfrac{q}{b}\left( d_{\hat{\beta}-2}{b_{{\pi _{\hat \alpha }}(\hat \beta  - 2)}} +  d_{\hat{\beta}}{b_{{\pi _{\hat \alpha }}(\hat \beta )}} \right)\left( {1 - 2{h_{{\pi _{\hat \alpha }}(\hat \beta  - 1)}}} \right) \\
&+ {p_{{\pi _{\hat \alpha }}(\hat \beta  - 1)}}\left( {1 - 2{h_{{\pi _{\hat \alpha }}(\hat \beta  - 1)}}} \right)\text{ \hspace{0.1cm} (mod {\em q}).} 
\end{aligned}
\end{equation}
\begin{equation}
 \begin{aligned}
&{c_{h,j}} - {c_{g,i}} - {c_{{h^s},j}} + {c_{{g^s},i}}\\
& = s\dfrac{q}{b}\left( {d_{\hat{\beta}-2}{a_{{\pi _{\hat \alpha }}(\hat \beta  - 2)}} + {d_{\hat{\beta}}a_{{\pi _{\hat \alpha }}(\hat \beta )}} - d_{\hat{\beta}-2}{b_{{\pi _{\hat \alpha }}(\hat \beta  - 2)}} - d_{\hat{\beta}}{b_{{\pi _{\hat \alpha }}(\hat \beta )}}} \right)\\
 &\equiv s\dfrac{q}{b}d_{\hat{\beta}}\left( {{a_{{\pi _{\hat \alpha }}(\hat \beta )}} - {b_{{\pi _{\hat \alpha }}(\hat \beta )}}} \right)      \equiv s\dfrac{q}{b}d_{\hat{\beta}}\text{ \hspace{0.1cm} (mod {\em q}).} 
\end{aligned}
\end{equation}

In (13), \(d_{\hat{\beta}}\) and \(b\) are prime each other, implying \(\sum\limits_{s = 0}^{b - 1}{\xi ^{{c_{h,j}} - {c_{g,i}}-{c_{{\rm{h^s}},j}} +{c_{g^s,i}}}} = 0\). Similarly, we can also obtain 
\begin{equation}
    \sum\limits_{s = 0}^{b - 1}{\xi ^{{c_{h,j}} - {c_{g,i}}-{c_{{\rm{h^s}},j}} +{c_{g^s,i}}}} = 0
\end{equation}
If \(m < {\pi _{\hat \alpha }}\left( {\hat {\beta} \; - 1} \right) \le n + m\), note that \(a_{\pi \left( {t - 1} \right)}^s = i_{\pi \left( {t - 1} \right) - n}^s\) and \({a_{\pi \left( {t - 1} \right)}} = {i_{\pi \left( {t - 1} \right) - n}}\) according to (9). By following the similar argument as mentioned above, we can get
\begin{equation}
\sum\limits_{s = 0}^{b - 1}{\xi ^{{c_{h,j}} - {c_{g,i}}-{c_{{\rm{h}},j^s}} +{c_{g,i^s}}}} = 0
\end{equation}
By combining the two cases above, we conclude that the array set is a \(\left( {{{N}}^{k + 1}},{b^{m},{b^{n}}} \right)\) - GCAS

\appendix
\section*{Appendix B}
\section*{Proof of theorem 2}
\textit{Proof:} 
We set integers \(h = g + {u_1}\), \(j = i + {u_2}\). Let \(\left( {{g_1},{g_2}, \ldots ,{g_m}} \right)\), \(\left( {{h_1},{h_2}, \ldots ,{h_m}} \right)\), \(\left( {{i_1},{i_2}, \ldots ,{i_n}} \right)\) and \(\left( {{j_1},{j_2},\ldots ,{j_n}} \right)\) be the \(b_1\)-ary representation of integers \(g\), \(h\) and the \(b_2\)-ary representation of integers \(j\), \(i\), respectively. Set
a \(N_1\)-ary vector \(\bm{\mu}=[\mu_1,\mu_2,…,\mu_k]\) and a \(N_2\)-ary vector \(\bm{\mu^{\prime}}=[\mu_1^{\prime},\mu_2^{\prime},…,\mu_k^{\prime}]\).

\begin{multline}
\sum\limits_{c \in \mathcal{A}} {\sum\limits_{g = 0}^{{b_1^m} - 1} {\sum\limits_{i = 0}^{{b_2^n} - 1} {{\xi ^{{c_{g + {u_1},i + {u_2}}} - {c_{g,i}}}}} } }  \\
= \sum\limits_{g = 0}^{{b_1^m} - 1} {\sum\limits_{i = 0}^{{b_2^n} - 1} {\sum\limits_{\bm{\mu}}{\sum\limits_{\bm{\mu}^{\prime}}} {{\xi ^{{c_{g + {u_1},i + {u_2}}} - {c_{g,i}}}}} } }  = 0,
\end{multline}
To prove the \(\mathcal{A}\) is GCAS, for any \( c \in \mathcal{A}\), we need to demonstrate when \(\left( {{u_1},{u_2}} \right) \ne \left( {0,0} \right)\), equation (16) holds.  

The proof of Theorem 2 will be discussed in four cases in what follows. 

{\em Case 1:} \(u_1 > 0, u_2 \ge 0\) and \({g_{{\pi _\alpha }\left( 1 \right)}} \ne {h_{{\pi _\alpha }\left( 1 \right)}}\) or \({g_{k_1+1}} \ne {h_{k_1+1}}\). 

{\em Case 2:} \(u_1 > 0, u_2 \ge 0\) and \({g_{{\pi _\alpha }\left( 1 \right)}} = {h_{{\pi _\alpha }\left( 1 \right)}}\) and \({g_{k_1+1}} = {h_{k_1+1}}\). 

{\em Case 3:} \(u_1 = 0, u_2 \ge 0\) and \({j_{{\sigma _\mu }\left( 1 \right)}} \ne {i_{{\sigma _\mu }\left( 1 \right)}}\) or \({j_{k_1+1}} \ne {i_{k_1+1}}\).

{\em Case 4:} \(u_1 = 0, u_2 > 0\) and \({j_{{\sigma _\mu }\left( 1 \right)}} = {i_{{\sigma _\mu }\left( 1 \right)}}\) and \({j_{k_1+1}} = {i_{k_1+1}}\).  

{\em Case 1} and {\em Case 3} are similar to {\em Case 1} in Theorem 1, and {\em Case 2} and {\em Case 4} are similar to {\em Case 2} in Theorem 1. The proof is omitted here to keep it brief.
\bibliographystyle{IEEEtran}
\small\bibliography{references}
\end{document}